\documentclass{aa}

\usepackage{graphicx}
\usepackage{txfonts}
\newcommand{\va}{v_{\mathrm{A}}}
\newcommand{\vae}{v_{\mathrm{A,e}}}
\newcommand{\vai}{v_{\mathrm{A,i}}}

\newcommand{\der}{{\rm d}}

\newcommand{\rhoi}{\rho_{\rm i}}
\newcommand{\rhoe}{\rho_{\rm e}}
\newcommand{\xii}{\mbox{\boldmath{$\xi$}}}

\newcommand{\rhotr}{\rho_{\rm tr}}

\newcommand{\ui}{U_{\rm i}}
\newcommand{\ue}{U_{\rm e}}
\newcommand{\utr}{U_{\rm tr}}

\newcommand{\ra}{r_{\mathrm{A}}}

\begin{document}

   \title{Transverse waves in coronal flux tubes with thick boundaries: The effect of longitudinal flows}

   \author{Roberto Soler\inst{1,2}}

   \institute{Departament de F\'isica, Universitat de les Illes Balears, E-07122 Palma de Mallorca, Spain.    
   \and
  Institut d'Aplicacions Computacionals de Codi Comunitari (IAC$^3$), Universitat de les Illes Balears, E-07122 Palma de Mallorca, Spain.  \\
\email{roberto.soler@uib.es} }
   \date{Received XXX; accepted XXX}

 
  \abstract
   {Observations show that transverse magnetohydrodynamic (MHD) waves and flows are often simultaneously present in magnetic loops of the solar corona. The waves are resonantly damped in the Alfv\'en continuum because of  plasma and/or magnetic field nonuniformity across the loop. The resonant damping is relevant in the context of coronal heating, since it provides a mechanism to cascade energy down to the dissipative scales. It has been theoretically shown that the presence of flow affects the waves propagation and damping, but most of the studies rely on the unjustified assumption that  the transverse nonuniformity is confined to a boundary layer much thinner than the radius of the loop. Here we present a semi-analytic technique to explore the effect of flow on resonant MHD waves in coronal flux tubes with thick nonuniform boundaries. We extend  a published method, which was originally developed for a static plasma, in order to incorporate the effect of flow. We allowed the flow velocity to continuously vary within the nonuniform boundary from the internal velocity to the external velocity.  The analytic part of the method is based on expressing the wave perturbations in the thick nonunform boundary of the loop as a Frobenius series that contains a singular term accounting for the Alfv\'en resonance, while the numerical part of the method  consists of solving iteratively the transcendental dispersion relation together with the equation for the  Alfv\'en resonance position. As an application of this method, we investigated the impact of flow on the phase velocity and resonant damping length of MHD kink waves. With the present method, we consistently recover results in the thin boundary approximation obtained in previous studies. We have   extended those results to the case of thick boundaries. We also  explored the error associated with the use of the thin boundary approximation beyond its regime of applicability.}

   \keywords{Magnetohydrodynamics (MHD) --- Sun: atmosphere --- Sun: corona --- Sun: oscillations --- Waves}

   \maketitle
%

\section{Introduction}

High-resolution observations have shown the ubiquitous presence of nearly incompressible transverse waves propagating in the solar corona \citep[see, e.g.,][]{Tomczyk2007,Tomczyk2009,McIntosh2011,Morton2016}. From the theoretical point of view, these waves are interpreted as magnetohydrodynamic (MHD) kink waves in magnetic flux tubes \citep[see, e.g.,][]{erdelyi2007,VanDoorsselaere2008,Mathioudakis2013,jess2015}. In solar coronal conditions, long-wavelength MHD kink waves  are almost incompressible and their dominant restoring force is magnetic tension \citep{goossens2012}. Due to plasma density and/or magnetic field nonuniformity across the waveguide,  MHD kink waves undergo the process of resonant absorption in the Alfv\'en continuum, by which their energy is transferred to localized  Alfv\'en waves that later develop small scales because of phase mixing \citep[see, e.g.,][]{lee1986,poedts1989,pascoe2012,Goossens2014,soler2015}. This process naturally brings wave energy down to dissipative scales at which it can be thermalized and heat the coronal plasma \citep[see, e.g., ][]{ionson1978}.

In addition to waves, observations often show the presence of flows along coronal loops \citep[see the review by][]{reale2014}. Most of the measured Doppler velocities associated to flows are typically in the range $\sim 5-30$~km~s$^{-1}$ \citep[e.g.,][]{delzanna2008,Winebarger2013}, although velocities of $\sim 50$~km~s$^{-1}$ have also been reported frequently \citep[e.g.,][]{Brekke1997,Doschek2008}, and even larger velocities up to $\sim 72-123$~km~s$^{-1}$ have been observed \citep{ofman2008}. Considering that the expected value of the Alfv\'en velocity in the solar corona is $\sim 1000$~km~s$^{-1}$, these observed flow speeds correspond to small fractions of the coronal  Alfv\'en velocity. Observations of flows with Alfv\'enic speeds  ($\gtrsim 500-1000$~km~s$^{-1}$) are scarce and often related to very energetic or explosive events \citep[e.g.,][]{innes2003,nitta2012}.

The presence of flows along the coronal waveguides modifies the behavior and properties of the MHD waves \citep[see, e.g.,][to name a few works]{naka1995,terra2003,ofman2008,ruderman2010}. In the case of resonantly damped kink waves, the effect of flow has been studied in a number of papers. \citet{goossens1992} presented an analytic theory based on the use of jump conditions for the wave perturbations at the Alfv\'en resonance.  \citet{goossens1992} derived an approximate expression for the damping rate of the waves, which was applicable to the case that the  variation of density and flow velocity across the waveguide were strictly confined to a thin layer, in other words, the so-called thin boundary approximation.  \citet{terradas2010} and \citet{soler2011} use the analytic formalism of  \citet{goossens1992} to study the effect of flow on the resonant damping of standing and propagating kink waves in coronal loops.   \citet{terradas2010} and \citet{soler2011}  also complement their results with the use of fully numerical resistive eigenvalue computations, which show a good agreement with the analytic approximations.

The thin boundary approximation is useful from an analytic point of view, because it allows the derivation of a simple expression for the damping rate. However, there  is no observational justification for the use of such an approximation. Indeed, some observations indicate that coronal loops are largely inhomogeneous in the transverse direction  \citep[see, e.g.,][]{Aschwanden2003,goddard2017}, and numerical simulations point out that  the waves themselves may contribute to generating wide nonuniform layers owing to nonlinear Kelvin–Helmholtz instabilites \citep[see][]{Goddard2018}. Therefore, it is important to explore the properties of resonant kink waves beyond the thin boundary approximation. Moreover, the role of flows need to be explored beyond the limit of thin transitions.

In the absence of flow the resonant damping of kink waves in loops with thick boundaries is investigated by \citet{vandoorsselaere2004} and \citet{arregui2005} using numerical resistive MHD eigenvalue computations.  \citet{paperI} revisit the same problem  in ideal MHD, but with an entirely different semi-analytic approach. The analytic part of the method is based on expressing the wave perturbations in the thick nonuniform boundary of the waveguide as a Frobenius series that contains a singular term accounting for the Alfv\'en resonance, while the numerical part of the method simply consists of solving a transcendental equation that plays the role of the dispersion relation. Since the technique of  \citet{paperI} is much faster than the solution of the resistive eigenvalue problem, detailed parameter studies can be tackled. The solutions provided by the method of  \citet{paperI} have been successfully used in a number of  papers: \citet{paperII} test the error associated with the use of the thin boundary approximation  beyond its theoretical range of applicability; \citet{Goossens2014}  discuss the kink quasi-mode displacement field in a tube with a wide boundary; \citet{arregui2015b} perform Bayesian inference, model comparison, and model-averaging techniques to infer the cross-field density structuring in coronal waveguides; \citet{soler2015} compare the solution provided by the Frobenius-based method with the temporal evolution obtained by expressing the kink wave as a superposition of Alfv\'en continuum modes; and \citet{soler2017} investigates the behavior of fluting modes in transversely nonuniform tubes. Independently, a similar series expansion approach (but without a resonant term) is used to study sausage waves in nonuniform tubes \citep[see][]{guo2016}.

The purpose of the present work is to further extend the method  of \citet{paperI} by incorporating the effect of mass flow along the waveguide. Both the density and the flow velocity are allowed to vary across the flux tube in a nonuniform layer of arbitrary thickness. Section~\ref{sec:model} presents the mathematical formalism, which is largely based on that by \citet{paperI} but with the appropriate modifications  to incorporate the flow. One of the most important differences is that, because of the spatially varying flow velocity, the equation for the radial position of the Alfv\'en resonance becomes a transcendental equation that has to be  solved iteratively along with the dispersion relation.  In order to verify the method, approximate results in the thin tube, thin boundary, and slow flow approximations are obtained and compared with previous expressions in the literature. As an application of this method, we explored the effect of flow on the phase velocity and resonant damping length of forward and backward propagating kink waves in coronal tubes with thick boundaries, described in Section~\ref{sec:appli}. Finally, some concluding remarks and prospects for future studies are given in Section~\ref{sec:conclusions}.

\section{Method}
\label{sec:model}

\subsection{Background}

As the background configuration to represent a coronal waveguide we consider a straight magnetic cylinder of radius $R$ embedded in a uniform and unbounded plasma. Cylindrical coordinates were used, with $r$, $\varphi$, and $z$ denoting the radial, azimuthal, and longitudinal coordinates, respectively. The magnetic field is straight and constant and along the axis of the cylinder, namely ${\bf B} = B {\bf 1}_z$. The mass density, $\rho$, is  uniform in the azimuthal and longitudinal directions and nonuniform in the radial direction, so that $\rho = \rho(r)$. We considered the following radial dependence for the density,
\begin{equation}
 \rho(r) = \left\{
\begin{array}{lll}
\rhoi, & \textrm{if} & r \leq R - l/2, \\
\rhotr(r), & \textrm{if} & R -l/2 < r < R +  l/2,\\
\rhoe, & \textrm{if} & r \geq R+l/2,
\end{array}
\right.
\end{equation}
where $\rhoi$ and $\rhoe$ are internal and external constant densities and $\rhotr(r)$ is the continuous density profile that  connects the internal plasma to the external plasma. We considered $\rhoi > \rhoe$ to represent a tube that is denser than the surrounding plasma. The thickness of the nonuniform boundary layer, $l$, is arbitrary and can take any value between $l = 0$ (abrupt jump) and  $l = 2R$  (fully inhomogeneous tube).  

In  addition, we considered the presence of a field-aligned mass flow, namely ${\bf U} = U {\bf 1}_z$, where $U$ is the flow velocity. As in the case of the density, we assume the flow velocity to be  uniform in the azimuthal and longitudinal directions and nonuniform in the radial direction, so that $U = U(r)$. For simplicity, we assumed that the radial dependence of the flow velocity mimics that of the density, namely
\begin{equation}
 U(r) = \left\{
\begin{array}{lll}
\ui, & \textrm{if} & r \leq R - l/2, \\
\utr(r), & \textrm{if} & R -l/2 < r < R +  l/2,\\
\ue, & \textrm{if} & r \geq R+l/2,
\end{array}
\right.
\end{equation}
where $\ui$ and $\ue$ are internal and external flow velocities and $\utr(r)$ denotes the spatially-dependent flow velocity in the transitional layer.

\subsection{Linear perturbations}

Linear ideal MHD waves are superimposed on the background state.  To study coronal transverse waves, we considered the linearized ideal MHD equations in the $\beta=0$ approximation, where $\beta$ refers to the ratio of the thermal pressure to the magnetic pressure. Therefore, the basic equations used in the present work are
 \begin{eqnarray}
 \rho\left( \frac{\partial {\bf v}}{\partial t} + {\bf U} \cdot \nabla {\bf v} + {\bf v} \cdot \nabla {\bf U} \right) &=& \frac{1}{\mu} \left( \nabla \times {\bf b} \right) \times {\bf B}, \label{eq:mom}\\
\frac{\partial {\bf b}}{\partial t} - \nabla \times \left( {\bf U} \times {\bf b}  \right) &=& \nabla \times \left( {\bf v} \times {\bf B} \right), \label{eq:induc}
\end{eqnarray}
where ${\bf v} = (v_r,v_\varphi,v_z)$ is the velocity perturbation, ${\bf b} = (b_r,b_\varphi,b_z)$ is the magnetic field perturbation, and $\mu$ is the magnetic permeability. In addition, the plasma Lagrangian displacement, $\xii = (\xi_r,\xi_\varphi,\xi_z)$, is related to the velocity perturbation and background flow by
\begin{equation}
{\bf v} = \frac{\partial \xii}{\partial t}  +  \left({\bf U} \cdot \nabla \right)\xii  - \left( \xii \cdot \nabla \right) {\bf U}. \label{eq:xii}
\end{equation} 

From here on we adopt the so-called quasi-mode approach, which assumes that the nonuniform waveguide supports global  modes \citep[see, e.g.,][for a discussion on the validity of this approach]{Goossens2014}. We expressed the temporal dependence of perturbations as $\exp(-i \omega t)$, where $\omega$ the global mode frequency.  In addition, we Fourier-analyzed the perturbations along the uniform $\varphi$- and $z$-directions, so that the perturbations are put proportional to $\exp(i m \varphi + i k_z z)$, where $m$ and $k_z$ and the azimuthal and longitudinal wavenumbers, respectively. In the linear regime, different azimuthal and longitudinal wavenumbers do not interact with each other. Hence, we only retained the dependence of the perturbations on the radial direction. We find from Equation~(\ref{eq:xii}) that the relation between the components of the velocity perturbation and the Lagrangian displacement are
\begin{eqnarray}
v_r &=& - i  \Omega(r)\xi_r, \label{eq:vr} \\
v_\varphi &=& - i  \Omega(r)\xi_\varphi, \label{eq:vf} \\
v_z &=& - \frac{\partial U(r)}{\partial r} \xi_r, \label{eq:vz} 
\end{eqnarray}
where $\Omega(r)=\omega-k_z U(r)$ is the spatially-dependent Doppler-shifted frequency. We note that because of the presence of the spatially-dependent longitudinal flow, $v_z$ is nonzero even in the $\beta = 0$ approximation.

The following procedure closely follows that of \citet{paperI}, with the difference here we have included the effect of flow. We used the total pressure Eulerian perturbation, $P' = {\bf B}\cdot{\bf b}/\mu$, as our main variable. We combined Equations~(\ref{eq:mom}) and (\ref{eq:induc}) and, after some algebraic manipulations, we obtain a differential equation involving $P'$ alone, namely
\begin{eqnarray}
\frac{\partial^2 P'}{\partial r^2} &+& \left[ \frac{1}{r} - \frac{\frac{\der}{\der r} \left( \rho(r)\left( \Omega^2(r) - k_z^2 \va^2(r) \right) \right)}{\rho(r)\left( \Omega^2(r) - k_z^2 \va^2(r) \right)} \right] \frac{\partial P'}{\partial r} \nonumber \\
&+& \left( \frac{\rho(r)\left( \Omega^2(r) - k_z^2 \va^2(r) \right) }{B^2/\mu} - \frac{m^2}{r^2} \right) P' = 0, \label{eq:ptot}
\end{eqnarray}
where $\va^2(r) = B^2 /\mu \rho(r)$ is the square of the spatially-dependent Alfv\'en velocity. We note that Equation~(\ref{eq:ptot}) can also be obtained from Equation~(18) of \citet{goossens1992} in the $\beta=0$ case and in the absence of magnetic twist. 

The components of the Lagrangian displacement are related to $P'$ as
\begin{eqnarray}
\xi_r &=& \frac{1}{\rho(r)\left( \Omega^2(r) - k_z^2 \va^2(r) \right)}\frac{\partial P'}{\partial r}, \label{eq:xir} \\
\xi_\varphi &=& \frac{1}{\rho(r)\left( \Omega^2(r) - k_z^2 \va^2(r) \right)} \frac{im}{r} P'. \label{eq:xif} 
\end{eqnarray}
along with $\xi_z=0$ because of the $\beta = 0$ approximation. In the absence of flow, i.e., for $U(r) = 0$ so that $\Omega(r) = \omega$, Equations~(\ref{eq:ptot})--(\ref{eq:xif}) consistently revert to Equations~(4)--(6) of \citet{paperI}. 

\subsection{Solution in the internal and external plasmas}

In the regions with constant density and flow velocity,  Equation~(\ref{eq:ptot}) simplifies to
\begin{equation}
\frac{\der^2 P'}{\der r^2} + \frac{1}{r}  \frac{\der P'}{\der r} + \left( \frac{\Omega^2 - k_z^2\va^2}{\va^2} - \frac{m^2}{r^2} \right) P' = 0, \label{eq:bessel}
\end{equation}
where now  $\va$ and $\Omega$ are constant. Equation~(\ref{eq:bessel}) is the Bessel Equation and applies both in the internal ($r \leq R - l/2$) and external ($r \geq R + l/2$) plasmas. We use the subscripts `i' and `e' to denote quantities related to the internal and external plasmas, respectively.

In the internal plasma,  $P'$ must be regular at $r=0$. Thus, the physical solution of Equation~(\ref{eq:bessel}) is
\begin{equation}
P'_{\rm i} = A_{\rm i} J_{m}\left( k_{\perp,\rm i} r  \right), \label{eq:pin}
\end{equation}
where $A_{\rm i}$ is a constant, $J_m$ is the Bessel function of the first kind of order $m$, and
\begin{equation}
        k_{\perp,\rm i}^2 =  \frac{\Omega_{\rm i}^2 - k_z^2\vai^2}{\vai^2}.
\end{equation}
In the external plasma, we required that $P'$ vanishes when $r\to \infty$. This is the condition for the wave to be trapped. The physical solution to Equation~(\ref{eq:bessel}) is then
\begin{equation}
P'_{\rm e} = A_{\rm e} K_{m}\left( k_{\perp,\rm e} r  \right), \label{eq:pex}
\end{equation}
where again $A_{\rm e}$ is a constant, $K_m$ is the modified Bessel function of the first kind of order $m$, and
\begin{equation}
        k_{\perp,\rm e}^2 =  - \frac{\Omega_{\rm e}^2 - k_z^2\vae^2}{\vae^2}.
\end{equation}

As in \citet{paperI} we have focussed on trapped waves and discard leaky waves from the present investigation. In the absence of flow, leaky waves have been investigated in detail by, for example, \citet{cally1986,cally2003} in the case of tubes with a piecewise constant density, and by, for example, \citet{stenuit1999,nakariakov2012,guo2016} in the case of transversely nonuniform tubes. As shown in \citet{terradas2010} and \citet{soler2011}, when the flow velocity surpasses a certain threshold  the trapped waves are forced to become leaky waves  because their frequency is located above the external cut-off frequency. In the case of kink waves, i.e., for $m=\pm 1$, the transition to the leaky regime occurs for an internal flow velocity that is super-Alfv\'enic. Since the vast majority of observed flows in coronal flux tubes are sub-Alfv\'enic \citep[see][]{reale2014}, for simplicity we  restricted our analysis to sub-Alfv\'enic flows and so avoid the possibility that the waves  become leaky. If leaky waves were to be considered, the solution in the external plasma (Equation~(\ref{eq:pex})) should be expressed using Hankel functions and  the condition of out-going waves should be enforced, that is, that there is no energy input from infinity \citep[see, e.g.,][for more details on the treatment of leaky modes]{stenuit1999,guo2016}. By restricting ourselves to sub-Alfv\'enic flows, we are also discarding the triggering of the Kelvin-Helmholtz instability due to velocity shear at the boundary of the tube. The Kelvin-Helmholtz instability in flux tubes with longitudinal shear flows  has been extensively investigated in the literature \citep[e.g.,][]{Holzwarth2007,ryutova2010,zhelyazkov2015} and is not the subject of the present study.

\subsection{Solution in the nonuniform boundary layer}

In the nonuniform transitional layer, and for $m\neq 0$, Equation~(\ref{eq:ptot}) is singular at the specific position, $r=\ra$, where the resonant condition $\Omega^2(\ra) = k_z^2 \va^2(\ra)$ is satisfied. Here,  $\ra$  denotes the Alfv\'en resonance position, which is a regular singular point of  Equation~(\ref{eq:ptot}).  We can expand the resonant condition to find an implicit equation whose solution is the resonant position, namely
\begin{equation}
\omega^2 - 2 k_z U(\ra) \omega + k_z^2 \left( U^2(\ra) - \va^2(\ra)\right) =0. \label{eq:ra}
\end{equation} 
For arbitrary density and flow profiles, Equation~(\ref{eq:ra}) has to be solved numerically to find $\ra$. An analytic expression of $\ra$ is only possible  for very specific and simple profiles. We note that Equation~(\ref{eq:ra}) depends upon $\omega$, which shall be obtained from the dispersion relation that, in turn, requires  $\ra$ to be known. So, Equation~(\ref{eq:ra}) must be solved iteratively  along with the dispersion relation.

Concerning the behavior of the waves, the presence of the resonance causes wave damping due to coupling between the global modes (quasi-modes) and the localized Alfv\'en waves around the resonance position. The temporal evolution of these coupled modes \citep[see, e.g.,][]{terradas2006,pascoe2010,soler2015} reveals that the energy of the global mode is ideally transferred to the Alfv\'en continuum modes in the nonuniform layer. These Alfv\'en continuum modes develop small length scales due to phase mixing. Finally, the small scales are dissipated by some non-ideal process such as, resistivity or viscosity. The present approach, which is based on ideal quasi-modes, allows us to study the first phase of this process, that is, the damping of the global mode, while the generation of small scales and their dissipation would requite other approaches beyond the aim of the present study.

As in \citet{paperI} (see also references therein), we used the method of Frobenius to express the solution to Equation~(\ref{eq:ptot}) as an infinite power series expansion  around the resonance position $r=\ra$. The method assumes that there is only one resonance. The existence of multiple  resonances is only possible when the density and flow profiles have very peculiar shapes that hardly represent the actual conditions in coronal flux tubes. For instance, various resonances may occur if the density within the transitional layer is not monotonic and increases and decreases following an oscillatory pattern, or when the density varies smoothly but the flow velocity varies very abruptly. This last case is analyzed in some detail by \citet{terradas2010}.

We rewrite Equation~(\ref{eq:ptot}) as
\begin{equation}
\left( r - \ra \right)^2 h(r) \frac{\partial^2 P'}{\partial r^2} + \left( r - \ra \right) p(r)  \frac{\partial P'}{\partial r} + q(r)  P' = 0, \label{eq:ptrfro}
\end{equation}
where the functions $h(r)$, $p(r)$, and $q(r)$ are defined as
\begin{eqnarray}
h(r) &=& r^2 f(r),\\
p(r) & =& r \left( r - \ra \right)  \left( f(r)- r \frac{\partial f(r)}{\partial r} \right), \\
q(r) &=&  \left( r - \ra \right)^2 \left( \frac{\mu}{B^2} r^2 f(r) - m^2 \right) f(r).
\end{eqnarray}
The functions $h(r)$, $p(r)$, and $q(r)$ take the same form as in \citet{paperI}. The difference resides in the expression of the function $f(r)$, which now contains the effect of flow and is given by
\begin{equation}
f(r) = \rho(r) \left( \Omega^2(r) - k_z^2 \va^2(r) \right) = \Omega^2(r)\rho(r) - k_z^2 \frac{B^2}{\mu}.
\end{equation}

We assumed that both the density and the flow velocity are analytic functions at $r=\ra$. We performed a Taylor series of $f(r)$ around the location of the resonance as 
\begin{equation}
f(r) = \sum_{k=0}^{\infty} f_{k}  \left( r - \ra \right)^k,
\end{equation}
with $f_0=0$ and
\begin{equation}
f_k = \sum_{n=0}^k \rho_n \sum_{l=0}^{k-n} \Omega_l \Omega_{k-n-l} \qquad \textrm{for} \qquad k \geq 1,
\end{equation}
where $\rho_0 = \rho(\ra)$, $\Omega_0 = \omega-k_z U(\ra)$, and
\begin{eqnarray}
\rho_{k} &=& \frac{1}{k!} \left. \frac{\der^k \rho}{\der r^k} \right|_{r=\ra}, \\
\Omega_{k} &=& \frac{1}{k!} \left. \frac{\der^k \Omega}{\der r^k} \right|_{r=\ra} =- \frac{k_z}{k!} \left. \frac{\der^k U}{\der r^k} \right|_{r=\ra},
\end{eqnarray}
when $k \geq 1$.

The general solution to Equation~(\ref{eq:ptrfro}) is 
\begin{equation}
P'_{\rm tr}(r) = A_0 P'_{1}(r) + S_0 P'_{2}(r), \label{eq:seriesgen}
\end{equation}
where the subscript `tr' denotes  the transitional layer, $A_0$ and $S_0$ are constants, and $P'_{1}(r)$ and $P'_{2}(r)$ are two linearly independent solutions. We find the two independent solutions with the help of a Frobenius series expansion around the regular singular point $r=\ra$. The indicial equation  is obtained from the coefficient of the lowest power of $\left(r-\ra\right)$, and it shows that zero and two are the two possible indices of the expansion.  Then, the two linearly independent solutions are
\begin{eqnarray}
P'_{1}(r) &=& \left(r-\ra\right)^2 \sum_{k=0}^{\infty} a_k \left(r-\ra\right)^k, \label{eq:regular} \\
P'_{2}(r) &=& \sum_{k=0}^{\infty} s_k \left(r-\ra\right)^k  + \mathcal{C} P'_{1}(r) \ln\left(r-\ra\right), \label{eq:singular}
\end{eqnarray}
where $\mathcal{C}$ is the coupling constant and $a_k$ and $s_k$ are the series coefficients.  We consider that the causal branch of the logarithm is the one where $\ln\left(r-\ra\right) = \ln\left(\ra-r\right) \pm i\pi$ if $r<\ra$, where the criterion for choosing either the $+$ sign or the $-$ sign is not arbitrary but based on the physical argument that the effect of the resonance is to produce the damping of the waves. The coupling constant, $\mathcal{C}$, is independent of the density and flow profiles and is given by 
\begin{equation}
\mathcal{C} =\frac{m^2}{2\ra^2}.
\end{equation}
For sausage waves, meaning when $m=0$, $\mathcal{C} = 0,$ the singular logarithmic term is dropped from  $P'_{2}(r)$. As a consequence, no resonant damping occurs if $m=0$. Conversely, the coefficients  $a_k$ and $s_k$ depend upon the choice of the density and flow profiles. General expressions of the coefficients $a_k$ and $s_k$  are given in the Appendix~\ref{app1}. 

In the simplified case that the nonuniform layer is  thin compared with the tube radius,  it suffices to keep  terms up to $\mathcal{O}(l/R)$ in the Frobenius series. This is the so-called thin boundary (TB) approximation. In that simple scenario, we find 
\begin{eqnarray}
P'_{\rm tr}(r) & \sim & S_0, \\
\xi_{r,\rm tr}(r) & \sim & \frac{S_0}{f_1}\frac{m^2}{\ra^2} \ln\left( r - \ra \right), \\
\xi_{\varphi,\rm tr}(r) & \sim & i \frac{S_0}{f_1}\frac{m}{\ra} \frac{1}{r-\ra}.
\end{eqnarray}
In a thin nonuniform layer the total pressure perturbation is constant, the radial  displacement jumps logarithmically, and the azimuthal displacement behaves as $1/(r-\ra)$. This behavior agrees with that found in previous works \citep[e.g.,][]{goossens1992}. In the present work we have used the full Frobenious solution that contains this fundamental behavior as well as the corrections beyond the TB approximation owing to the larger thickness of the nonuniform layer. 

\subsection{Dispersion relation} 

The conditions that the total pressure perturbation, $P'$, and the radial component of the Lagrangian displacement, $\xi_r$, are continuous at  $r=R - l/2$ and $r=R + l/2$ provide us with a system of four algebraic equations for the constants $A_{\rm i}$, $A_{\rm e}$, $A_0$, and $S_0$. The dispersion relation of the waves is then obtained from the requirement that there is a nontrivial solution of the system, in other words, by setting the associated determinant equal to zero. The expression of the general dispersion relation is
\begin{eqnarray}
&& \frac{\frac{k_{\perp,\rm e}}{\rhoe \left( \Omega_{\rm e}^2 - k_z^2\vae^2 \right)} \frac{K'_{m}\left[ k_{\perp,\rm e}(R+l/2)\right]}{K_{m}\left[ k_{\perp,\rm e}(R+l/2)\right]}\mathcal{G}_+ -  \Xi_+}{\frac{k_{\perp,\rm e}}{\rhoe \left( \Omega_{\rm e}^2 - k_z^2\vae^2 \right)} \frac{K'_{m}\left[ k_{\perp,\rm e}(R+l/2)\right]}{K_{m}\left[ k_{\perp,\rm e}(R+l/2)\right]}\mathcal{F}_+ -  \Gamma_+} \nonumber \\
&-&\frac{\frac{k_{\perp,\rm i}}{\rhoi \left( \Omega_{\rm i}^2 - k_z^2\vai^2 \right)} \frac{J'_{m}\left[ k_{\perp,\rm i}(R-l/2)\right]}{J_{m}\left[ k_{\perp,\rm i}(R-l/2)\right]}\mathcal{G}_- - \Xi_-}{\frac{k_{\perp,\rm i}}{\rhoi \left( \Omega_{\rm i}^2 - k_z^2\vai^2 \right)} \frac{J'_{m}\left[ k_{\perp,\rm i}(R-l/2)\right]}{J_{m}\left[ k_{\perp,\rm i}(R-l/2)\right]}\mathcal{F}_- -  \Gamma_-} = 0. \label{eq:reldisper}
\end{eqnarray}
The quantities $\mathcal{G}_\pm$, $\mathcal{F}_\pm$, $\Xi_\pm$, and $\Gamma_\pm$ are defined as
\begin{eqnarray}
\mathcal{G}_\pm & = & \sum_{k=0}^\infty a_k \zeta_\pm^{k+2}, \\
\mathcal{F}_\pm & = & \sum_{k=0}^\infty \left( s_k \zeta_\pm^{k} + \frac{m^2}{2\ra^2}\ln\left( \zeta_\pm \right) a_k \zeta_\pm^{k+2} \right), \\
\Xi_\pm  & = &  \frac{1}{\sum_{k=0}^\infty f_{k+1}\zeta_\pm^k}\sum_{k=0}^\infty (k+2)a_k \zeta_\pm^{k}, \\\
\Gamma_\pm & = & \frac{1}{\sum_{k=0}^\infty f_{k+1}\zeta_\pm^k}\sum_{k=0}^\infty\left(  k s_k \zeta_\pm^{k-2} + \frac{m^2}{2\ra^2} a_k \zeta_\pm^{k} \right. \nonumber \\
 &&+ \left. \frac{m^2}{2\ra^2} \ln\left( \zeta_\pm\right)  (k+2)a_k \zeta_\pm^{k}\right),
\end{eqnarray} 
with $\zeta_\pm = R \pm \frac{l}{2} -\ra$. In the absence of flow the dispersion relation reverts to that given in \citet{paperI}. 

Equation~(\ref{eq:reldisper}) is a transcendental equation with roots that are to be found numerically. To do this, we used a numerical routine based on that used by \citet{paperI}. A difference of this case from the one without flow in  \citet{paperI} is that the resonance position, $\ra$, also needs to be found numerically from Equation~(\ref{eq:ra}). The transverse density profiles used by  \citet{paperI} allow analytic expressions for $\ra$, which is not possible here because of the presence of flow. Hence, the present method iteratively solves Equations~(\ref{eq:ra}) and (\ref{eq:reldisper}) until both solutions converge. We note that the dispersion relation involves series with infinite number of terms. To proceed numerically we must truncate the infinite series so that only the first $N$ terms are accounted for. To make sure that the number of terms considered is large enough for the error to be negligible, we  performed convergence tests by increasing $N$ until a good convergence of the solution is obtained. Typically, we consider $N=51$.

We shall  find the solutions to Equation~(\ref{eq:reldisper}) in the case of fixed, real, and positive $k_z$. So, we impose a particular wavelength of the perturbations, namely $\lambda = 2\pi/k_z$. Then, the solution  is a complex frequency, namely $\omega = \omega_{\rm R} + i \omega_{\rm I}$, where the subscripts R and I denote the real and imaginary parts, respectively.  The sign of $\omega_{\rm R}$ indicates the direction of wave propagation. If $\omega_{\rm R} > 0$ the wave propagates toward the positive $z$-direction (forward propagation), while propagation is in the opposite direction when $\omega_{\rm R} < 0$ (backward propagation). The phase velocity is computed as 
\begin{equation}
v_{ph} = \frac{\omega_{\rm R}}{k_z},
\end{equation}
where the sign of $v_{ph}$ follows the same rules as the sign of $\omega_{\rm R}$. On the other hand, $\omega_{\rm I}$ is related to the damping rate of the waves, so that $\omega_{\rm I} < 0$ for both directions of propagation.  If the waves are not overdamped, i.e., if $\left|\omega_{\rm I} \right| < \left| \omega_{\rm R} \right|$, we can define  the exponential damping length of the waves, $L_{\rm D}$, as the distance the waves need to travel for their amplitude to be reduced by a factor of  $e$, namely \citep[see][]{Tagger1995}
\begin{equation}
L_{\rm D} = \left| \frac{1}{\omega_{\rm I}} \frac{\partial \omega_{\rm R}}{\partial k_{z}}  \right|.
\end{equation}
The damping length so defined is positive for both forward and backward propagating waves.

\subsection{Recovering the thin tube, thin boundary, and slow flow approximation}

Equation~(\ref{eq:reldisper})  is valid for arbitrary values of $l/R$ and $k_z R$.  An appropriate way to check the validity of Equation~(\ref{eq:reldisper}) is to recover the dispersion relation previously obtained in the literature in the thin tube (TT, $k_z R \ll 1$) and thin boundary (TB, $l/R \ll 1$) approximations \citep[e.g.,][]{goossens1992,terradas2010,soler2011}. To do so, we  kept terms up to $\mathcal{O}(l/R)$ in the expressions of  $\mathcal{G}_\pm$, $\mathcal{F}_\pm$,   $\Xi_\pm$, and $\Gamma_\pm$, and  approximate  $\ra\approx R$ as consistent with the assumption that the boundary layer is so thin that the resonance position is necessarily close to $r=R$.  Then, we find  $\mathcal{G}_\pm \approx 0$, $\mathcal{F}_\pm  \approx  1$, $\Xi_\pm \approx 2/f_1$, and
\begin{eqnarray}
\Gamma_+ & \approx & \frac{m^2/R^2}{f_1}\left(\ln  \frac{l}{2}  + \frac{1}{2} \right), \\
\Gamma_- & \approx & \frac{m^2/R^2}{f_1}\left(\ln \frac{l}{2} \pm i\pi + \frac{1}{2} \right).
\end{eqnarray} 
In order to make further analytic progress, we assumed that the magnetic tube is thin and  use the first-order expansion for small arguments and $m\neq 0$ of  the Bessel functions in  Equation~(\ref{eq:reldisper}) \citep[see, e.g.,][]{abra} and also take  $R-l/2\approx R + l/2 \approx R$, namely
\begin{eqnarray}
\frac{J'_{m}\left( k_{\perp,\rm i}\left( R - l/2 \right)\right)}{J_{m}\left( k_{\perp,\rm i}\left( R - l/2 \right)\right)} & \approx & \frac{J'_{m}\left( k_{\perp,\rm i}R\right)}{J_{m}\left( k_{\perp,\rm i}R\right)} \approx \frac{m}{k_{\perp,\rm i}R}, \\
 \frac{K'_{m}\left( k_{\perp,\rm e}\left( R + l/2 \right)\right)}{K_{m}\left( k_{\perp,\rm e}\left( R + l/2 \right)\right)} & \approx &  \frac{K'_{m}\left( k_{\perp,\rm e}R\right)}{K_{m}\left( k_{\perp,\rm e}R\right)} \approx - \frac{m}{k_{\perp,\rm e}R}. 
\end{eqnarray}
After long but straightforward algebraic manipulations, the TT and TB approximation of  Equation~(\ref{eq:reldisper}) can be cast as
\begin{eqnarray}
&&\rhoi \left( \Omega_{\rm i}^2 - k_z^2\vai^2 \right) + \rhoe \left( \Omega_{\rm e}^2 - k_z^2\vae^2 \right)  \nonumber \\
&&= i \pi \frac{m/R}{\left| f_1 \right|} \rhoi \left( \Omega_{\rm i}^2 - k_z^2\vai^2 \right) \rhoe \left( \Omega_{\rm e}^2 - k_z^2\vae^2 \right), \label{eq:tttbreldisper}
\end{eqnarray}
with
\begin{equation}
\left| f_1 \right|= \left| \Omega^2 \frac{\partial \rho}{\partial r} - 2 k_z \Omega \rho  \frac{\partial U}{\partial r} \right|_{r=R}.
\end{equation}
Equation~(\ref{eq:tttbreldisper})  agrees with Equation~(74) of \citet{goossens1992}, which is also used in \citet{terradas2010} and \citet{soler2011}.   Thus,  the TT and TB dispersion relation is correctly recovered from the more general Equation~(\ref{eq:reldisper}). We refer readers to those works to study the solutions of the TT and TB dispersion relation.

\begin{figure}
   \centering
\includegraphics[width=\columnwidth]{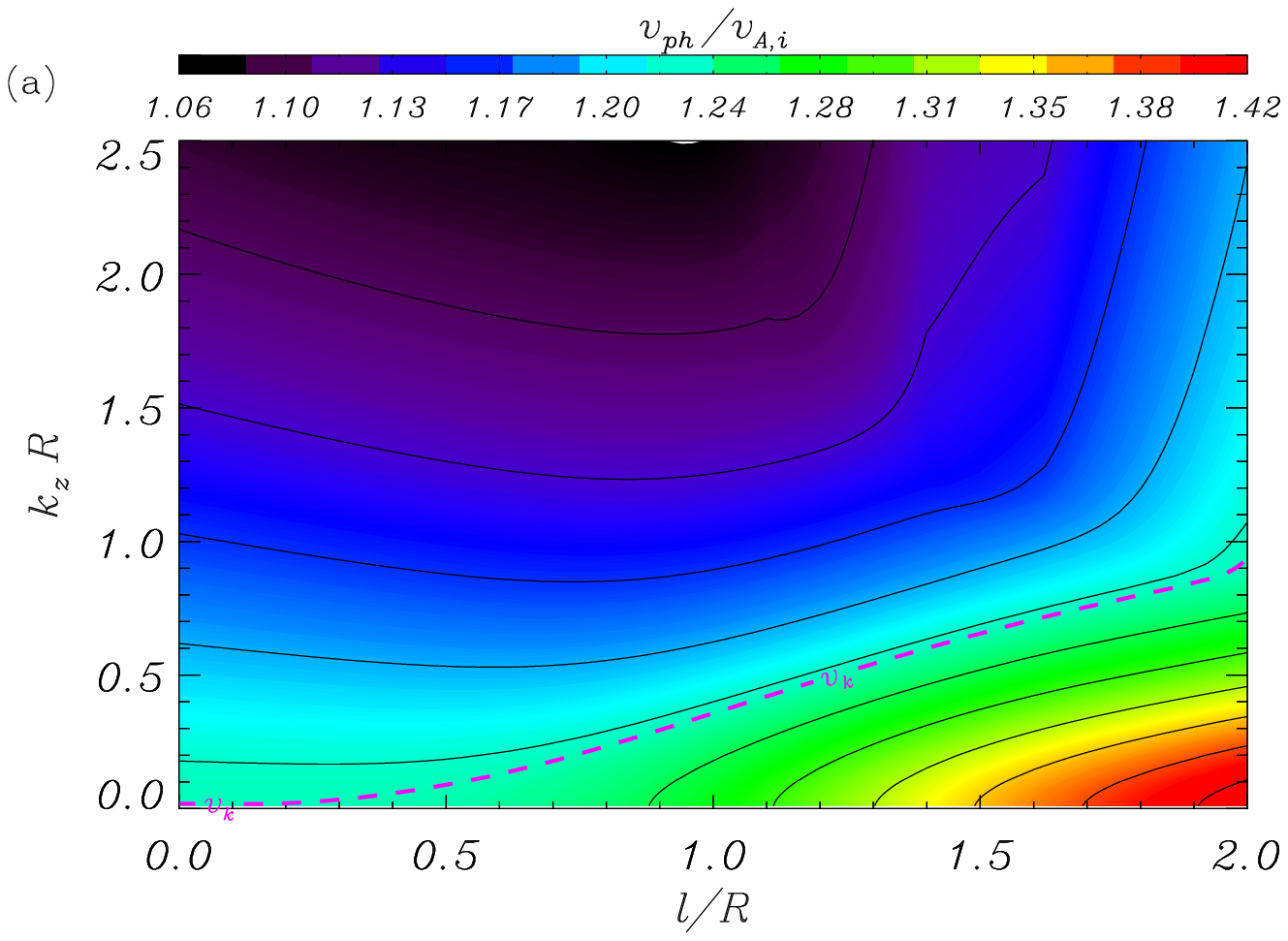} 
\includegraphics[width=\columnwidth]{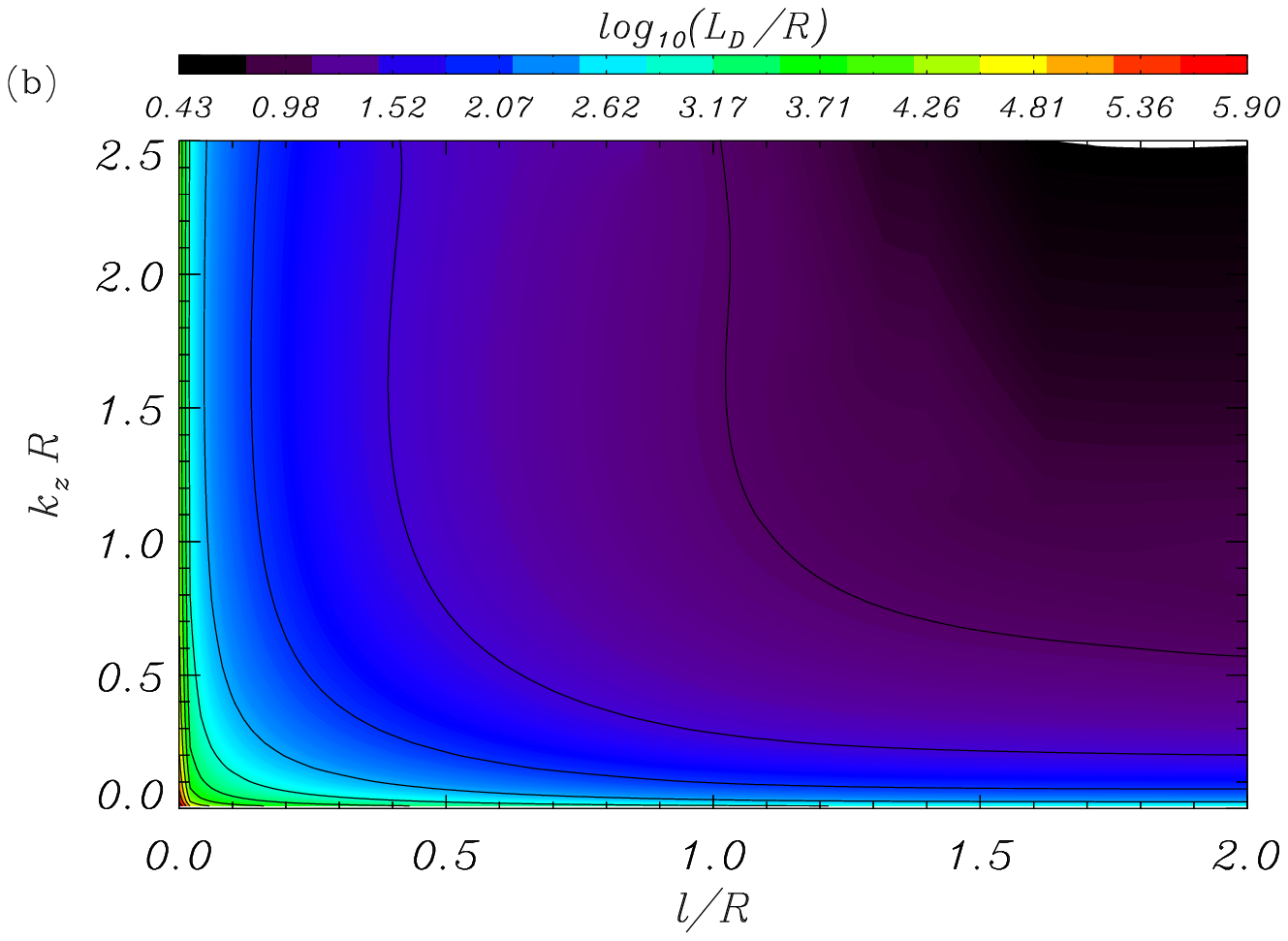}
      \caption{Results in the absence of flow. Contour plots of (a) $v_{ph}/\vai$ and (b)  $L_{\rm D}/R$ as functions of $l/R$ and $k_z R$. We note that $L_{\rm D}/R$ is given in logarithmic scale. The dashed purple line in panel (a) denotes $v_{ph} = v_k$.}
         \label{fig:noflow}
   \end{figure}

   \begin{figure}
   \centering
\includegraphics[width=\columnwidth]{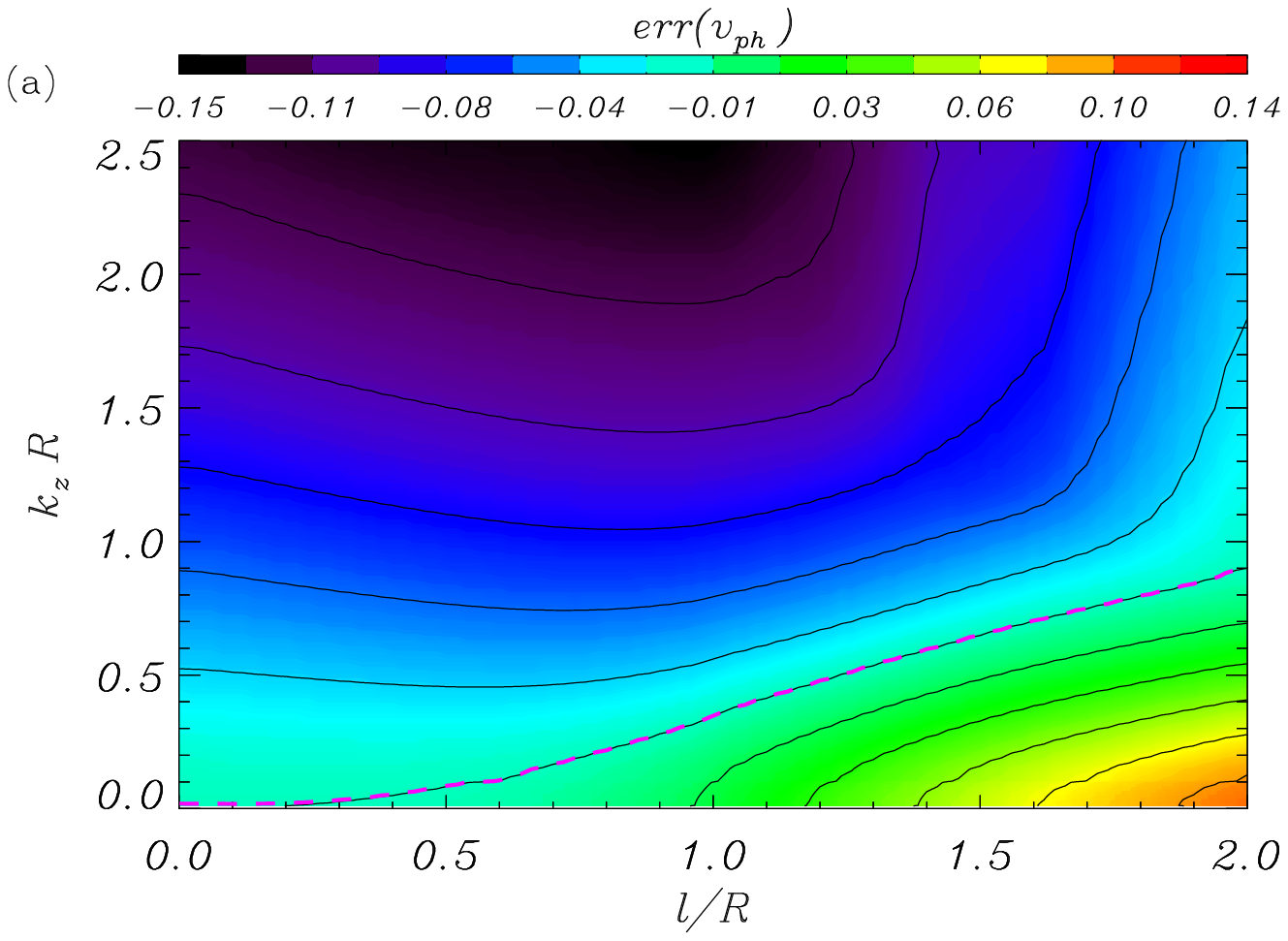} 
\includegraphics[width=\columnwidth]{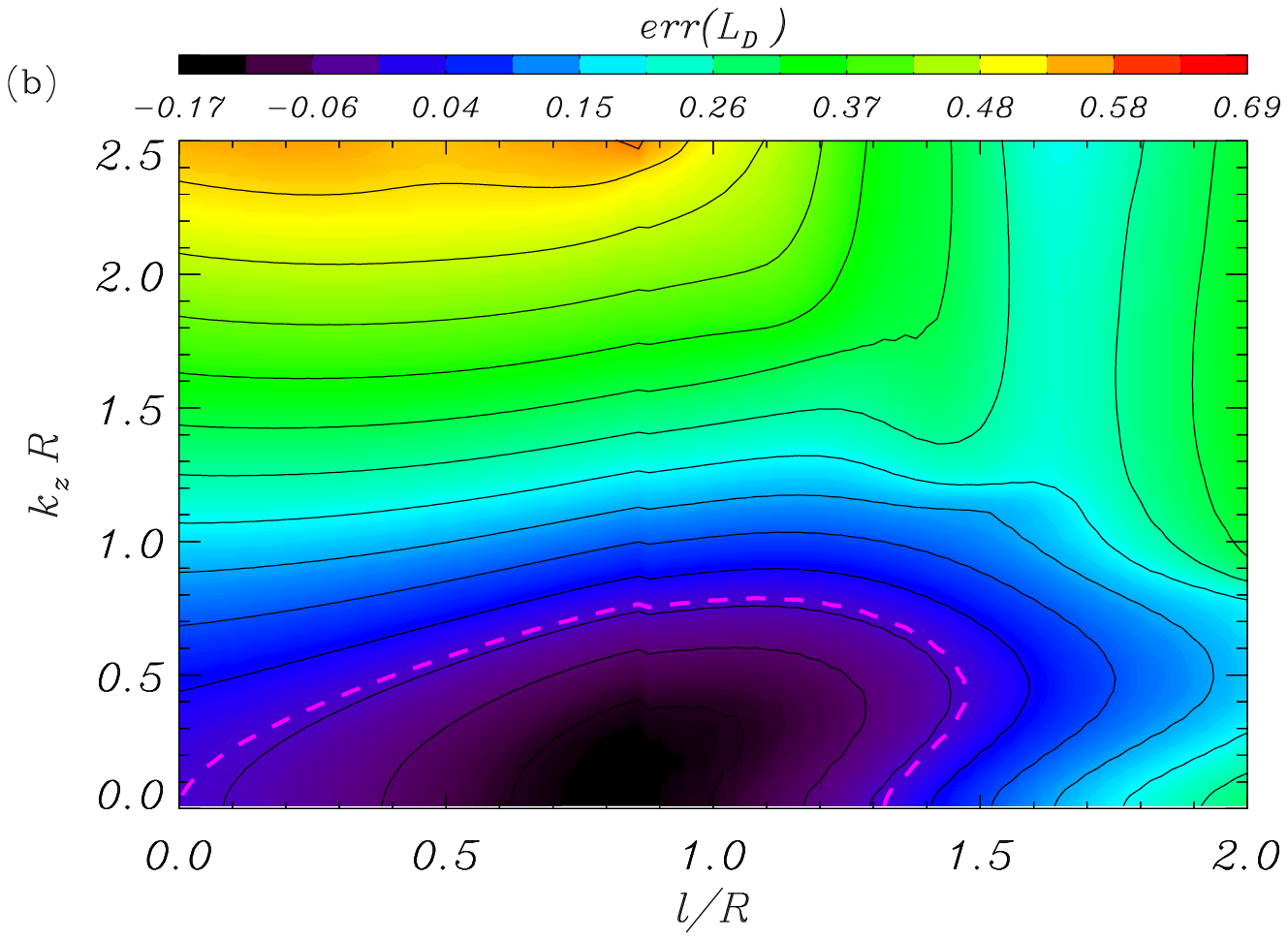}
      \caption{Results in the absence of flow. Contour plots of the errors associated with the approximate Equations~(\ref{eq:vphtttb}) and (\ref{eq:ldtttb}) for (a) $v_{ph}$ and (b)  $L_{\rm D}$ as functions of $l/R$ and $k_z R$. The dashed purple line in both panels denotes the contour of zero error.}
         \label{fig:noflowerr}
   \end{figure}

\citet{soler2011} considered Equation~(\ref{eq:tttbreldisper}) and obtained approximate expressions of $v_{ph}$ and $L_{\rm D} $ with $\ue = 0$ and in the case of slow flow (SF, $\ui/\vai\ll 1$), namely
\begin{eqnarray}
v_{ph} &\approx&  \pm v_k + \frac{\rhoi}{\rhoi+\rhoe}\ui, \label{eq:vphtttb} \\
L_{\rm D}  &\approx& \frac{2\pi}{k_z} \frac{F}{m} \frac{R}{l} \frac{\rhoi + \rhoe}{\rhoi - \rhoe} \frac{v_k}{v_{ph}}\left( 1 \pm \frac{2\rhoi}{\rhoi-\rhoe} \frac{\ui}{v_k} \right) \nonumber \\
& \approx & \frac{2\pi}{k_z} \frac{F}{m} \frac{R}{l} \frac{\rhoi + \rhoe}{\rhoi - \rhoe} \left( 1 \pm \sqrt{\frac{\rhoi}{2\left( \rhoi+\rhoe \right)}} \frac{\rhoi + 3 \rhoe}{  \rhoi-\rhoe } \frac{\ui}{\vai} \right), \label{eq:ldtttb}
\end{eqnarray}
where the $+$ and $-$ signs stand for forward and backward propagating waves respectively, $F$ is a numerical factor that depends on the shape of the transitional layer \citep[see][]{paperII}, and $v_k$ is the kink velocity given by
\begin{equation}
v_k = \sqrt{\frac{\rhoi\vai^2+\rhoe\vae^2}{\rhoi+\rhoe}} = \sqrt{\frac{2\rhoi}{\rhoi + \rhoe}} \vai.
\end{equation}
Although Equations~(\ref{eq:vphtttb}) and (\ref{eq:ldtttb}) are only strictly valid in the limits $k_zR \ll 1$, $l/R \ll 1$, and $\ui/\vai\ll 1$ they provide useful information regarding the effects of the various parameters. We note that  Equations~(\ref{eq:vphtttb}) and (\ref{eq:ldtttb}) predict that the effect of flow is to produce a linear correction in the flow velocity to both $v_{ph}$ and $L_{\rm D} $. We compare these approximations with the solutions of the general Equation~(\ref{eq:reldisper}), which remains valid beyond the range of applicability of Equations~(\ref{eq:vphtttb}) and (\ref{eq:ldtttb}). 

\section{Application}
\label{sec:appli}

\begin{figure*}
   \centering
\includegraphics[width=0.99\columnwidth]{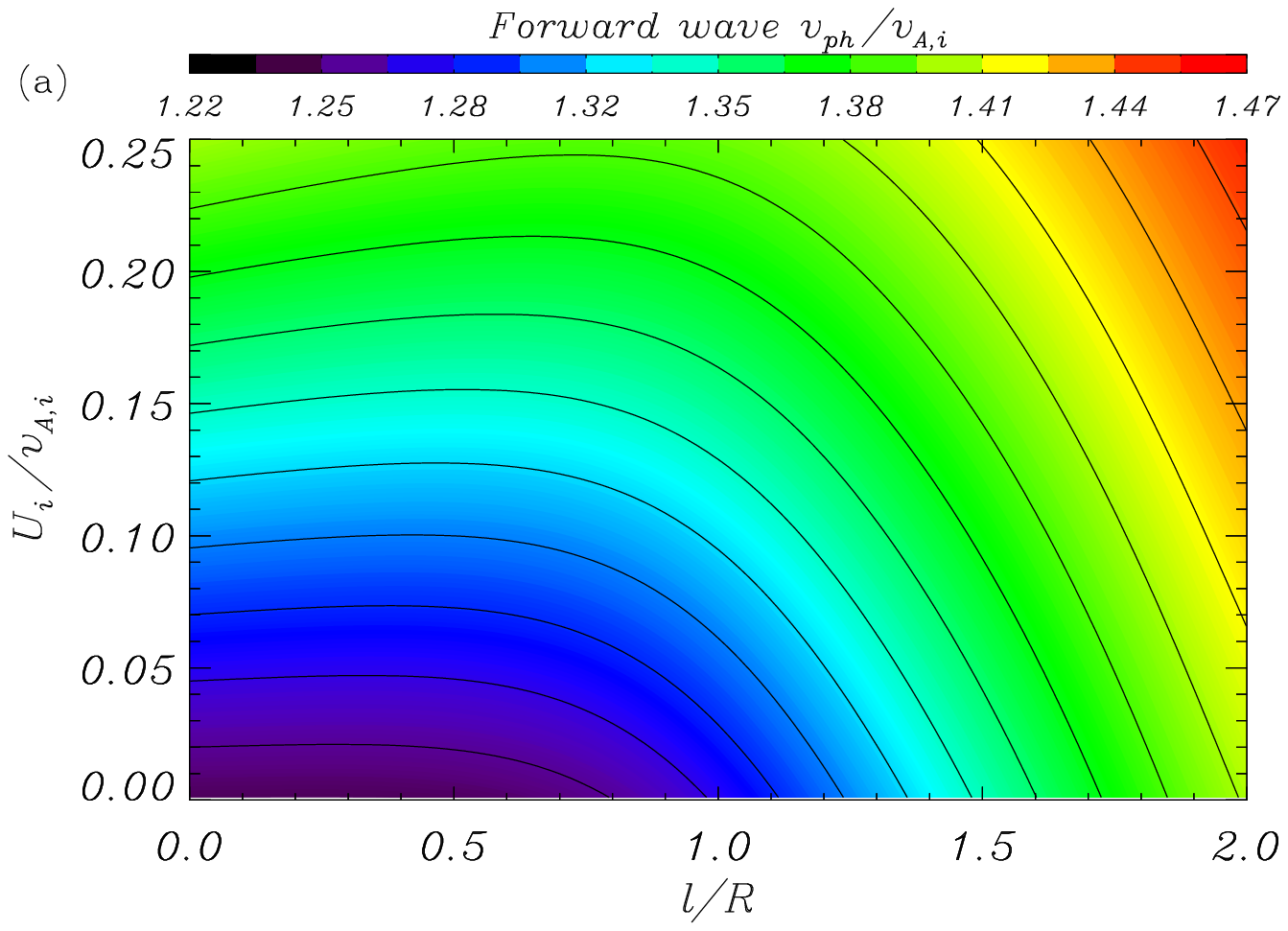} \includegraphics[width=0.99\columnwidth]{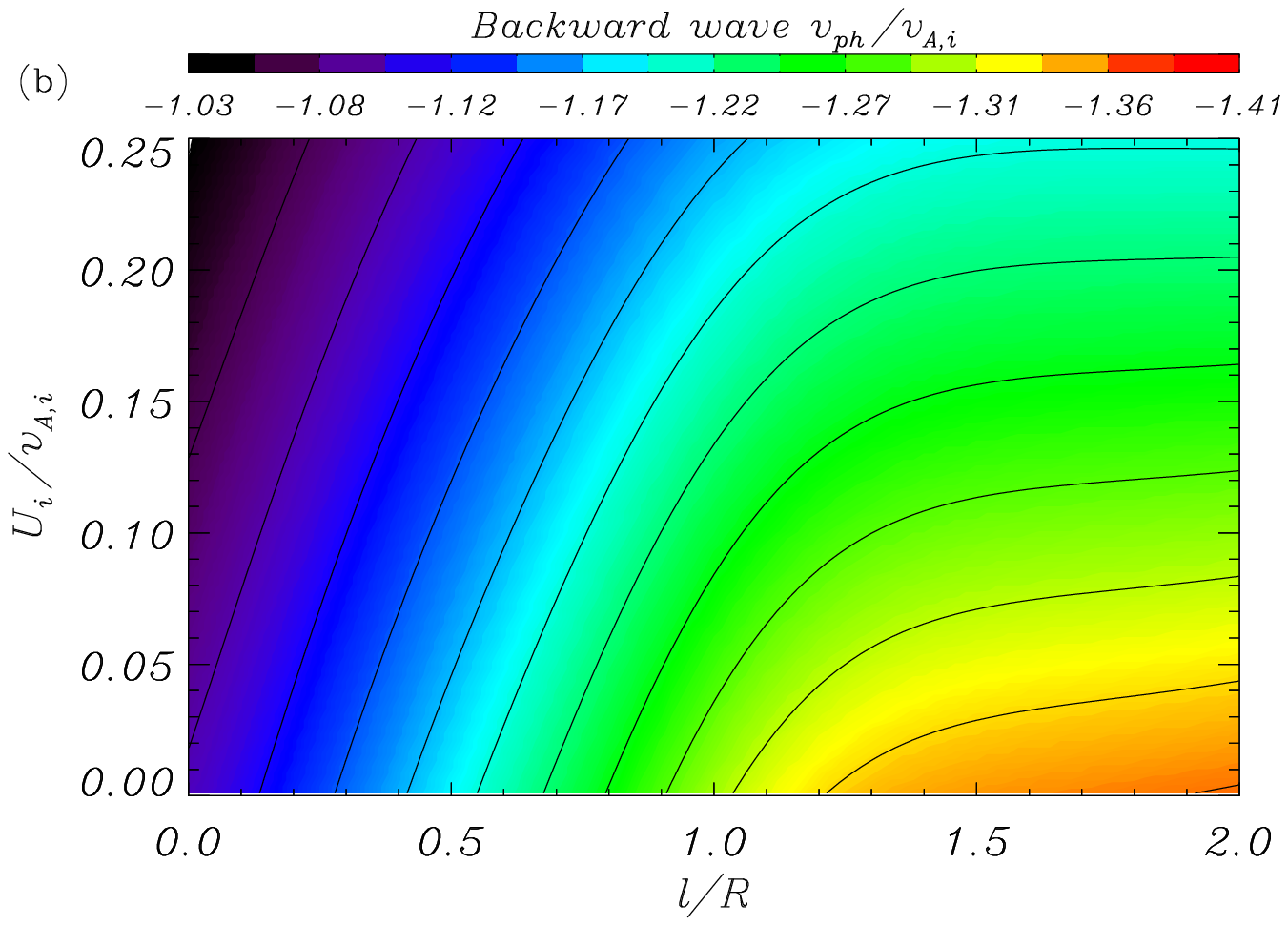} \\
\includegraphics[width=0.99\columnwidth]{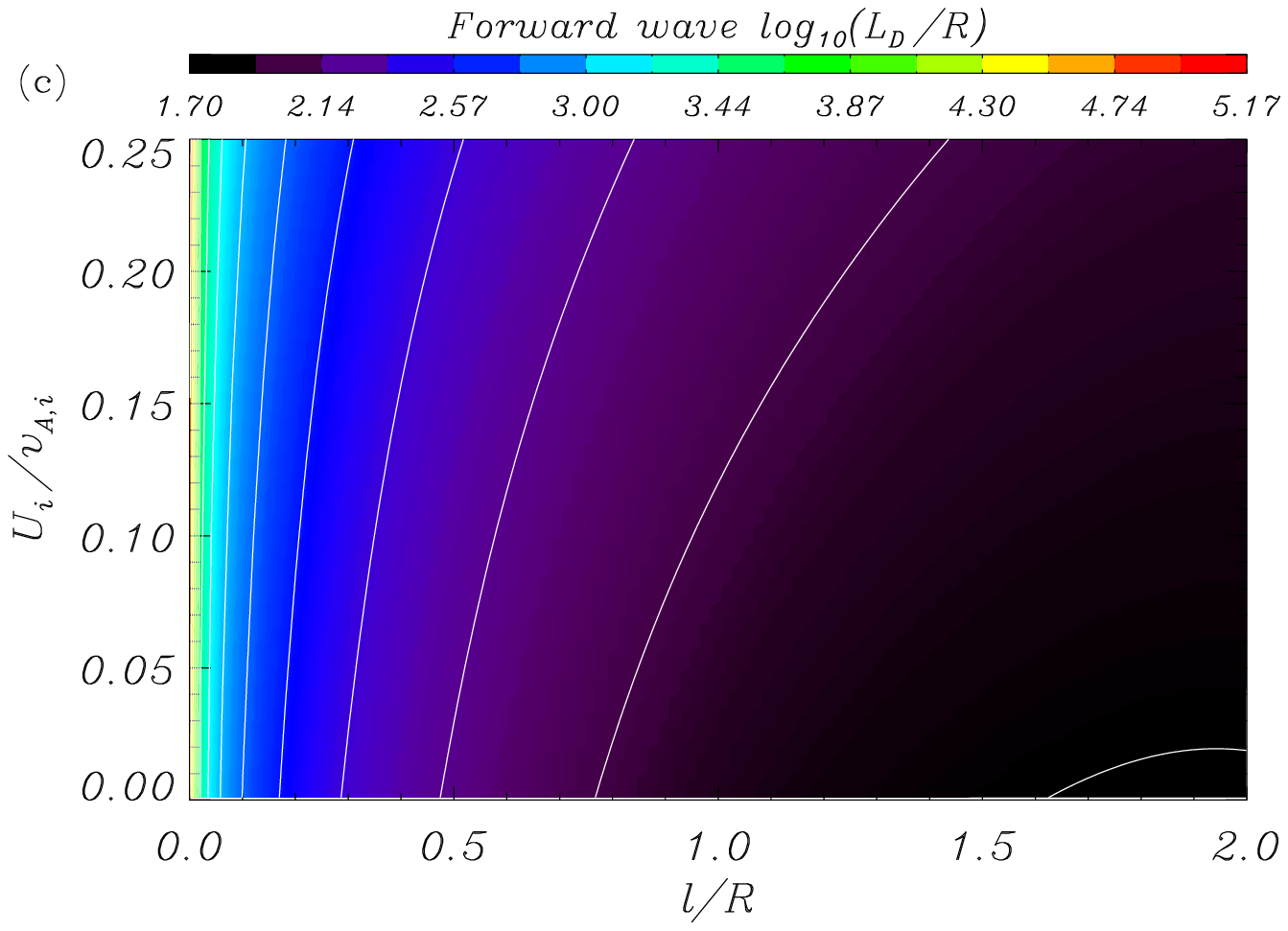} \includegraphics[width=0.99\columnwidth]{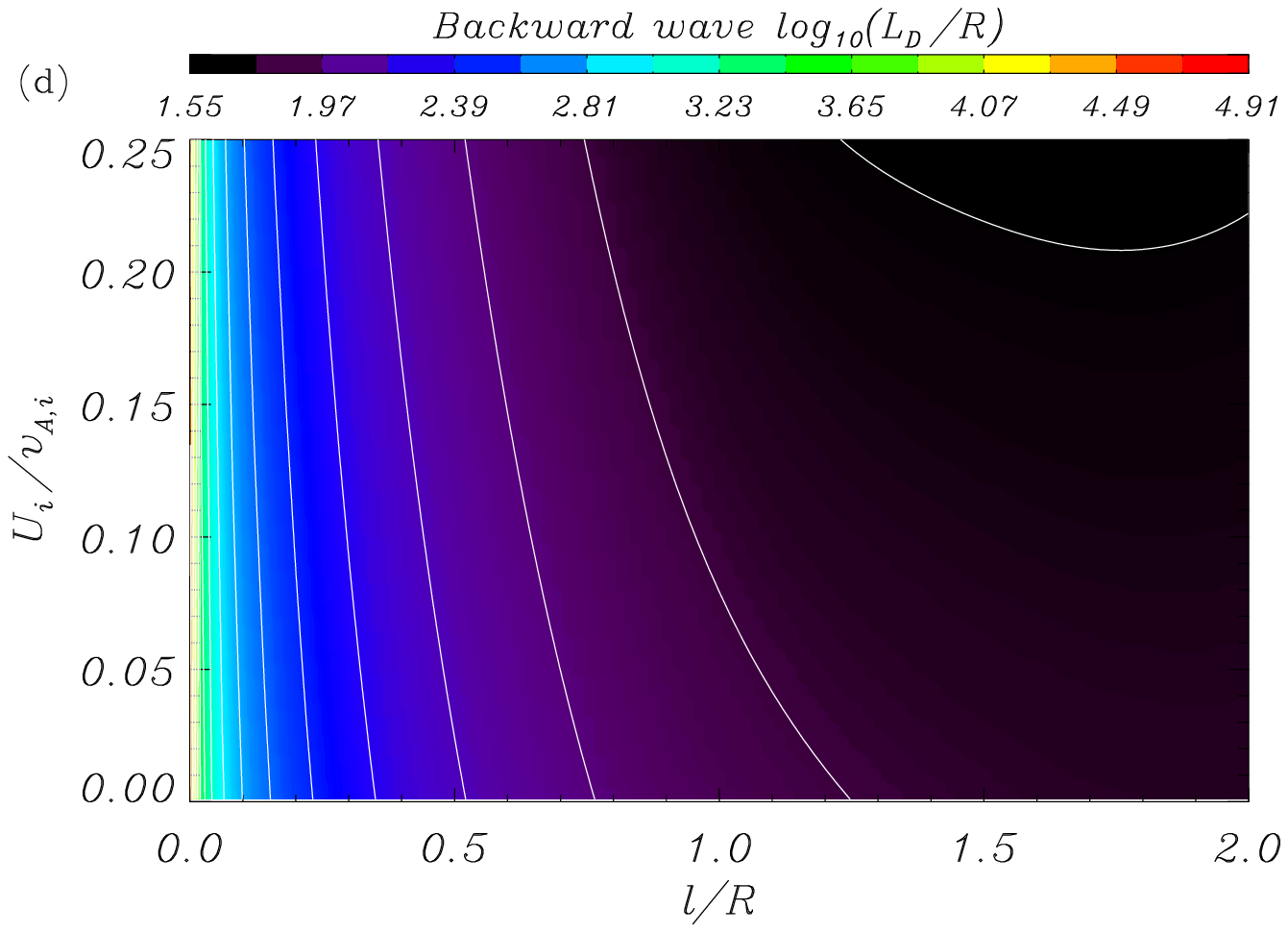} 
      \caption{Results in the presence of flow. Contour plots of $v_{ph}/\vai$ (panels (a) and (b)) and $L_{\rm D}/R$ (panels (c) and (d)) as functions of $l/R$ and $\ui/\vai$. Panels (a) and (c) are for the forward propagating wave, while panels (b) and (d) are for the backward propagating wave.   We note that $L_{\rm D}/R$ is given in logarithmic scale. We have used  $k_z R = 0.1$.}
         \label{fig:flowkz01}
   \end{figure*}

\begin{figure*}
   \centering
\includegraphics[width=0.99\columnwidth]{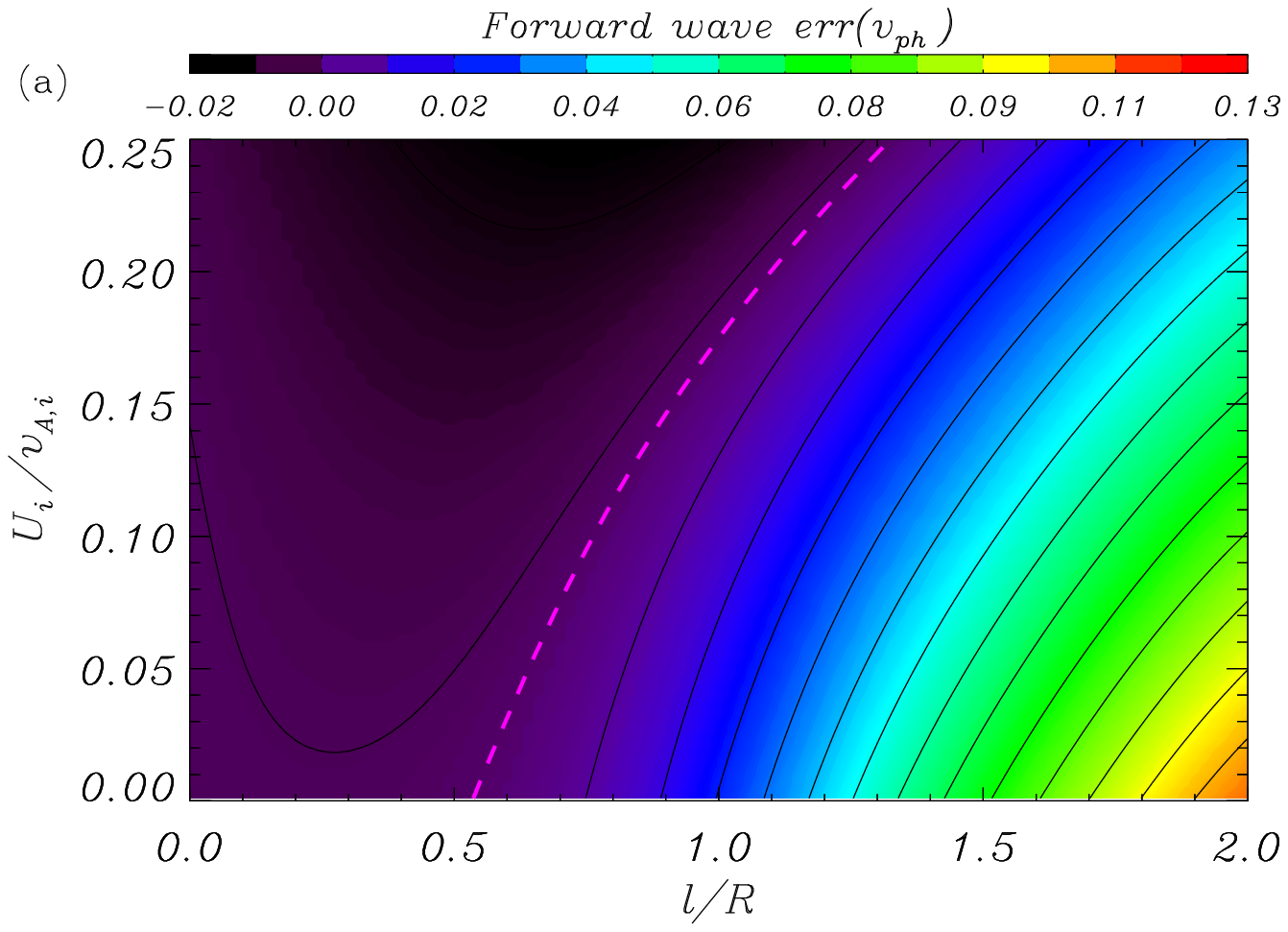} \includegraphics[width=0.99\columnwidth]{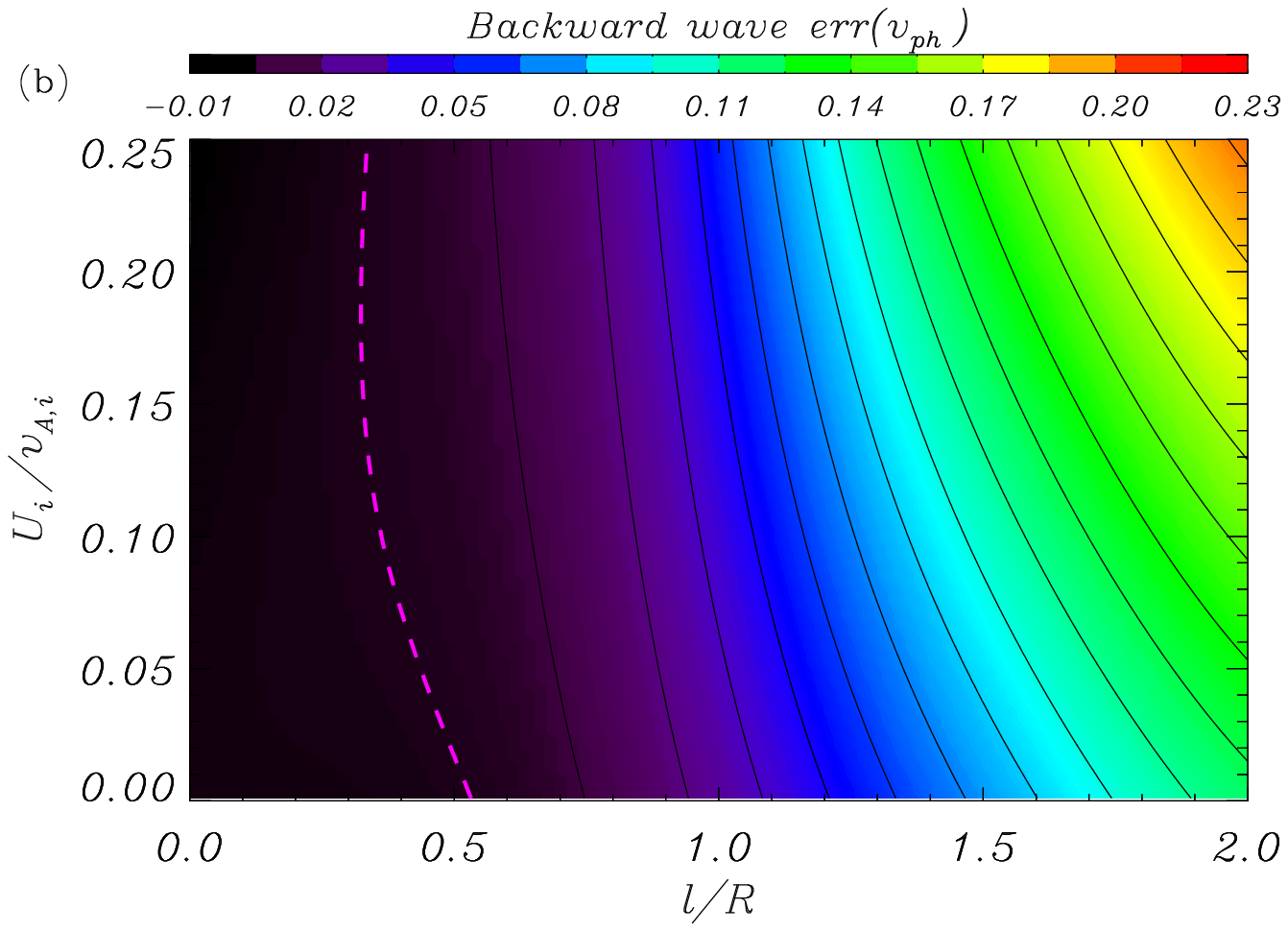} \\
\includegraphics[width=0.99\columnwidth]{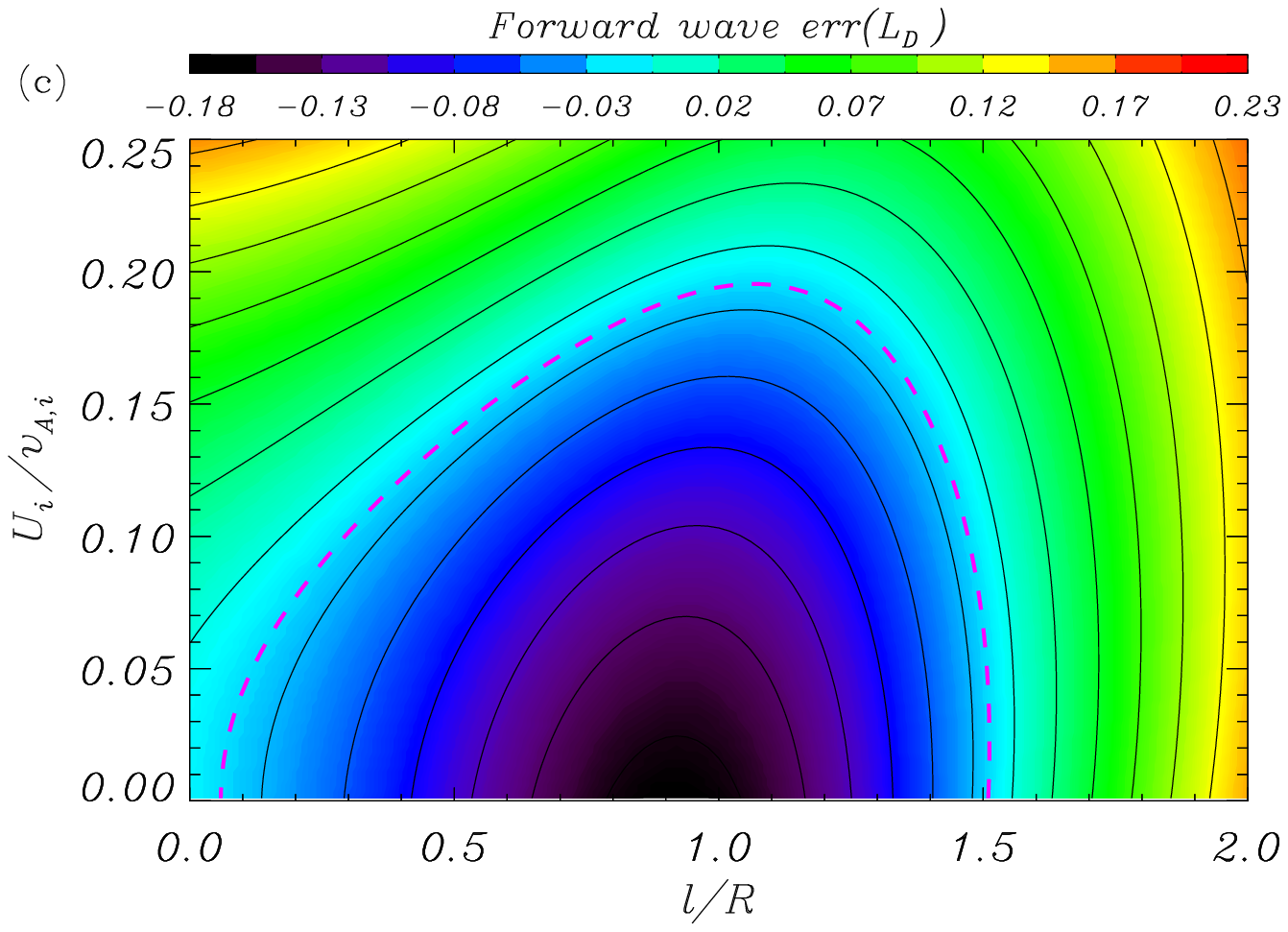} \includegraphics[width=0.99\columnwidth]{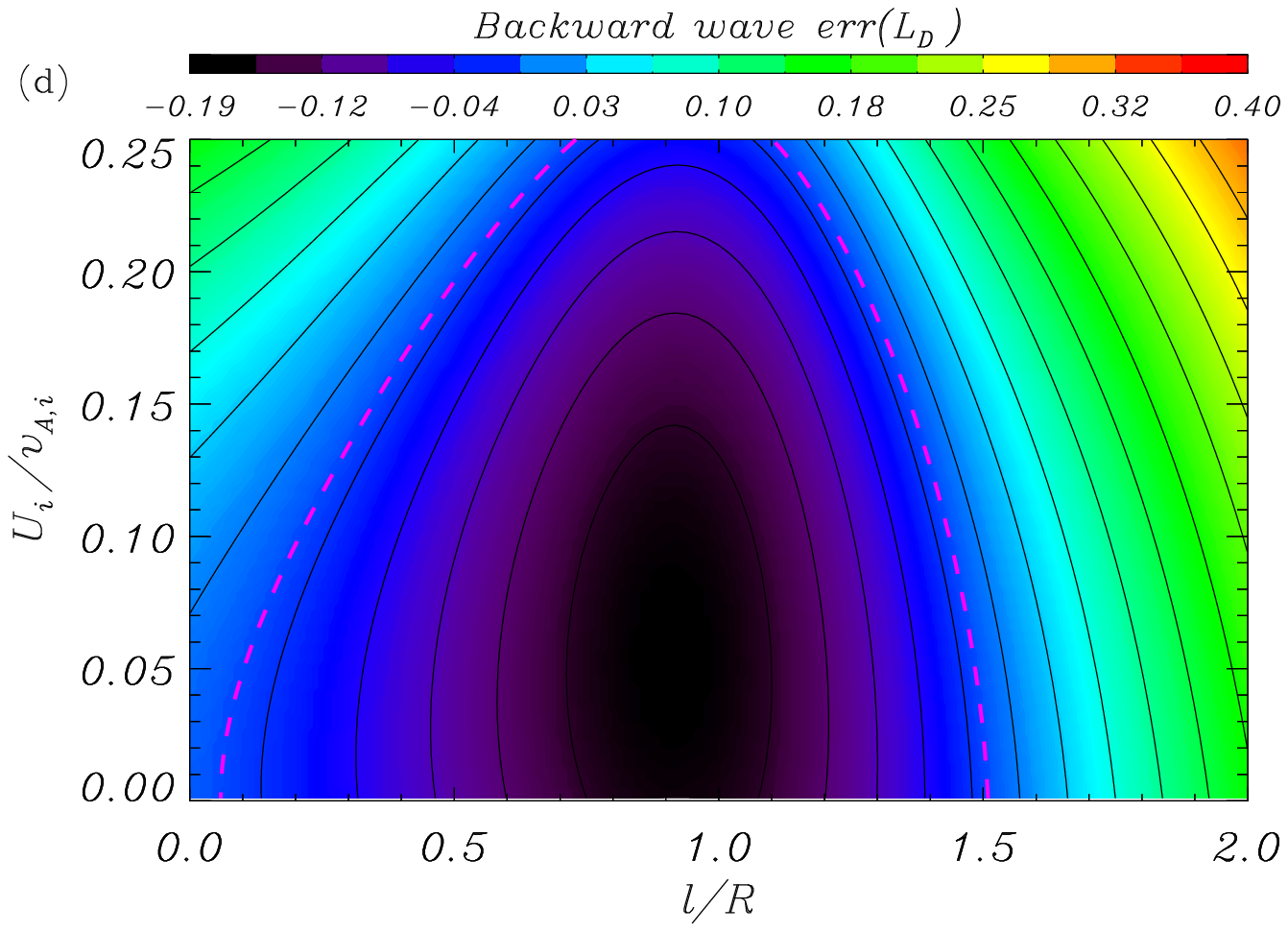} 
      \caption{Results in the presence of flow. Contour plots of the errors associated with the approximate Equations~(\ref{eq:vphtttb}) and (\ref{eq:ldtttb}) for $v_{ph}$ (panels (a) and (b)) and $L_{\rm D}$ (panels (c) and (d)) as functions of $l/R$ and $\ui/\vai$. Panels (a) and (c) are for the forward propagating wave, while panels (b) and (d) are for the backward propagating wave.   The dashed purple line  denotes the contour of zero error. We have used  $k_z R = 0.1$.}
         \label{fig:flowkz01err}
   \end{figure*}

As a  brief application of the method described above, we  computed results corresponding to the fundamental radial harmonic of the $m=1$ waves, in other words, the so-called kink mode. We note, however, that the dispersion relation is valid for any value of $m$ and for any radial harmonic as long as the modes are not leaky. In all the numerical solutions, we considered a sinusoidal transition for both the density and the flow velocity within the nonuniform boundary layer. For simplicity, we chose the reference frame so that the external plasma is static, meaning that we set $\ue = 0$. In addition, we set the ratio of the internal density to the external density to $\rhoi/\rhoe = 3$, which is representative of a coronal loop. For this density contrast, $v_k  \approx 1.22 \vai$. 

We investigated the behavior of $v_{ph}$ and $L_{\rm D}$ as functions of three dimensionless quantities: the ratio of the internal flow velocity to the internal Alfv\'en velocity, $\ui / \vai$, the relative thickness of the boundary layer, $l/R$, and dimensionless longitudinal wavenumber, $k_z R$. Typical values of $\vai$ and $R$ in coronal loops are $\vai \sim$~1,000~km~s$^{-1}$ and $R\sim$~3,000~km.

\subsection{Case without flow}

To start with, we considered the case with $\ui=0$. In this scenario, forward and backward propagating waves are degenerate, so for simplicity, only the forward wave was considered. This corresponds to the situation studied by \citet{paperI}. First of all, we check that the solutions given in \citet{paperI} are correctly recovered with the present numerical routine. We have fully  confirmed this for some selected configurations (this analysis is not shown here). 

We then focussed on investigating the dependence on $k_z R$, which was not done in \citet{paperI}. Here we have studied the impact of this parameter. Figure~\ref{fig:noflow} shows surface plots of $v_{ph}/\vai$ and $L_{\rm D}/R$ as functions of $l/R$ and $k_z R$. Concerning the phase velocity, we find that regardless the value of $k_z R$, there is an increasing trend in the phase velocity as $l/R$ increases, in agreement with the results of \citet{paperI} for fixed $k_z R$. Conversely, the effect of increasing $k_z R$ is the opposite, in other words,  the phase velocity decreases. Therefore, the parameters  $l/R$ and $k_z R$ have competing effects on the phase velocity. When $l=0$, the phase velocity  approaches the internal Alfv\'en velocity when $k_z R$ increases \citep[see][]{edwin1983}, but when $l\neq 0$ the phase velocity tends to a value somewhat larger than $\vai$ when $k_z R$ increases. This effect is due to the transverse nonuniformity. 

On the other hand,  $l/R$ and $k_z R$ have the same effect on the damping length, since $L_{\rm D}/R$ decreases when any of the two parameters increases. This result agrees qualitatively with the approximate analytic dependence of Equation~(\ref{eq:ldtttb}). From the physical point of view, the damping length becomes smaller when $l/R$ increases because the efficiency of resonant damping grows, while the effect of increasing $k_z R$ is to produce shorter wavelengths (i.e., higher frequencies) that are more efficiently damped.

In order to test the accuracy of the approximate Equations~(\ref{eq:vphtttb}) and (\ref{eq:ldtttb}), we  define the normalized errors of $v_{ph}$ and $L_{\rm D}$ associated with the use of the approximations  as
\begin{eqnarray}
{\rm err}\left( v_{ph} \right) &=& \frac{v_{ph,\rm Frob}-v_{ph,\rm app}}{v_{ph,\rm Frob}}, \\
{\rm err}\left( L_{\rm D} \right) &=& \frac{L_{\rm D, Frob}-L_{\rm D,app}}{L_{\rm D, Frob}},
\end{eqnarray}
where $v_{ph,\rm app}$ and $L_{\rm D,app}$ denote the analytic approximations given by Equations~(\ref{eq:vphtttb}) and (\ref{eq:ldtttb}), respectively, and $v_{ph,\rm Frob}$ and $L_{\rm D, Frob}$ are the actual results obtained with the present Frobenius-based method. Figure~\ref{fig:noflowerr} shows the normalized errors as functions of $l/R$ and $k_z R$. Very small errors are obtained when $l/R \ll 1$ and $k_z R \ll 1$, as consistent with the regime of applicability of the approximations. For larger values of the two parameters, the errors display a nonmonotonic behavior. In the case of the phase velocity, the largest errors are $\sim 15\%$ and are obtained either when $l/R \to 2$ and $k_z R \ll 1$ or when $l/R \ll 1$ and $k_z R \gg 1$. On the contrary, the error in the case of the damping length seems to be more dominated by $k_z R$. The largest values of ${\rm err}\left( L_{\rm D} \right)$ are $\sim 70\%$ and are obtained when $k_z R \gg 1$, while the error when  $k_z R \ll 1$ remains moderate even for large values of $l/R$. Remarkably, owing to the peculiar nonmonotonic behavior of the errors, we obtain zero errors of both $v_{ph}$ and $L_{\rm D}$ for some particular combination of parameters far beyond the regime of applicability of the approximations.

\subsection{Effect of flow}

We took flow into account and  exploited the new extension to the method. Figure~\ref{fig:flowkz01} shows surface plots of $v_{ph}/\vai$ and $L_{\rm D}/R$ as functions of $l/R$ and  $\ui / \vai$ for $k_z R = 0.1$. Because of the flow, different results are obtained for forward and backward waves, so Figure~\ref{fig:flowkz01} displays both. Physically, the effect of the flow is to drag the waves toward to direction of the flow, and hence a  shift of the phase velocity is produced. This results in larger phase velocities for the forward propagating wave and smaller phase velocities (in absolute value) for the backward propagating wave when compared with the case without flow. As in the absence flow, the effect of increasing $l/R$ is to increase the phase velocity (in absolute value) for both forward and backward waves. Thus, $l/R$ and  $\ui / \vai$ have the same effect on the phase velocity of forward waves, while these parameters have competing effects on the phase velocity of backward waves. 

Regarding the damping length, we find that for fixed $k_z R$, the behavior of $L_{\rm D}$ is essentially governed by $l/R$, while the role of the flow is to produce a small positive (for the forward wave) or negative (for the backward wave) shift with respect to the value for $\ui =0$. Hence, the backward wave is more efficiently damped than the forward wave, that is, the backward wave has a shorter $L_{\rm D}$. Qualitatively, this behavior agrees with the approximate analytic dependence of Equation~(\ref{eq:ldtttb}) and is also consistent with the resistive MHD results of \citet{soler2011}.

As before, we have computed the errors associated with the use of the approximations. These results are given in Figure~\ref{fig:flowkz01err}. In the case of the phase velocity, we obtain that the error grows with $l/R$ so that the largest errors are found when $l/R=2$. Clearly, the departure from the theoretical range of the TB approximation significantly reduces the accuracy of Equation~(\ref{eq:vphtttb}). In this regard, the error of backward wave phase velocity is larger than that of the forward wave. In the considered range of parameters, the maximum error of phase velocity is $\sim 13\%$ for the forward wave and $\sim 23\%$ for the backward wave. Conversely, the departure from the theoretical range of the SF approximation produces comparatively small errors. Equation~(\ref{eq:vphtttb})  predicts a linear shift of the phase velocity with the flow velocity, and such a dependence remains quite accurate in the whole range of considered flow velocities.

Next we considered to the error of the damping length. As in the case without flow, ${\rm err}(L_{\rm D})$ displays a nonmonotonic behavior with  $l/R$. This fact makes it is possible to find very small errors (even zero error) associated with the use of Equation~(\ref{eq:ldtttb}) when $l/R$ is large and some specific values of the flow velocity are considered. On the other hand, the dependence of ${\rm err}(L_{\rm D})$ with $\ui/\vai$ shows a much simpler relation, namely ${\rm err}(L_{\rm D})$ typically grows when $\ui/\vai$ increases. The largest values of ${\rm err}(L_{\rm D})$ plotted in Figure~\ref{fig:flowkz01err} are obtained when the flow velocity takes the maximum value used in the computations and, as before, the backward wave is the solution for which the approximation performs worst. As a general rule, the linear dependence on $\ui/\vai$ predicted by the analytic Equations~(\ref{eq:vphtttb}) and (\ref{eq:ldtttb}) as a consequence of adopting SF approximation is less accurate for $L_{\rm D}$ than for $v_{ph}$. 

\section{Concluding remarks}
\label{sec:conclusions}

The semi-analytic technique developed by \citet{paperI} to compute transverse waves in flux tubes with thick boundary layers has been extended  to incorporate the effect of longitudinal mass flows. The flow velocity is allowed to vary within the nonuniform boundary from the internal velocity to the external velocity. In the past, similar configurations have been studied analytically and/or numerically with  resistive eigenvalue computations but  in the case of tubes with thin transitions \citep[see, e.g.,][]{goossens1992,terradas2010,soler2011}. Two advantages of the present semi-analytic approach based on the Frobenius method are that nonuniform boundaries of arbitrary width can be considered and that its numerical implementation is much faster than resistive Eigenvalue computations. Therefore, detailed studies of the impact of the various model parameters on the wave properties are feasible with the present approach. We foresee future works that could exploit the method.

As an exemplary application, we have performed a parameter study of the effect of flow on the phase velocity and damping length of resonantly damped kink waves. First, we consistently recover the results in the TT, TB, and SF approximations obtained in previous works. Thanks to the present approach, we then extend those results beyond the range of applicability of the approximations. We have focussed on testing the validity of the TT, TB, and SF approximations against the actual results provided by the Frobenius-based method. While the TT and SF approximations perform relatively well for the considered wavenumbers and flow velocities, the use of the TB approximation implies a significant error when used beyond the limit $l/R \ll 1$. We note that we computed the results for a fixed value of the density ratio $\rhoi/\rhoe$ and boundary layer of sinusoidal shape, while \citet{paperII} showed that both ingredients also affect the accuracy of the approximations. Therefore, it is difficult to deduce a simple universal estimation of the error associated with the approximations, especially the TB approximation, since their accuracy is quite sensitive to the background configuration considered.

We have used the simplification that the density and the flow velocity follow that same radial dependence. \citet{terradas2010} and \citet{soler2011} considered the case that the density and flow velocity vary transversely to the tube within transitional layers of different width. They found that this has little impact on the results except in the case that the transitional layer for the flow velocity is much thinner than the corresponding layer for the density. In that case, the forward propagating wave damps faster than the backward propagating wave \citep[see details in][]{terradas2010,soler2011}. This is, however, a very peculiar situation that hardly represents the expected conditions in coronal flux tubes.

Besides the effect of flows, there are still various physical ingredients that could be incorporated to the semi-analytic Frobenius-based method. For instance, two additional improvements of the method would be to consider longitudinal stratification \citep[e.g.,][]{andries2005,arregui2005} and magnetic twist \citep[e.g.,][]{terradas2012,ruderman2015}. These extensions could be tackled in forthcoming works.

%
%

\begin{acknowledgements}
     RS acknowledges the support from grant AYA2017-85465-P (MINECO/AEI/FEDER, UE) and from the `Ministerio de Econom\'ia, Industria y Competitividad' and the `Conselleria d'Innovaci\'o, Recerca i Turisme del Govern Balear (Pla de ci\`encia, tecnologia, innovaci\'o i emprenedoria 2013-2017)' for the `Ram\'on y Cajal' grant RYC-2014-14970. 
\end{acknowledgements}

\bibliographystyle{aa} 
 \bibliography{refs.bib}

\begin{appendix}

\onecolumn

\section{Expressions of the Frobenius series coefficients}
\label{app1}

The expressions of the Frobenius series coefficients $a_k$ and $s_k$ are as follows:
\begin{eqnarray}
a_0 & = & 1, \\
a_1 & = & -\frac{2 f_1 - 2 f_2 \ra}{3 f_1\ra}a_0, \\
a_2 & = & - \frac{9 f_1 \ra a_1 + \left( 2 f_1 - 2f_2 \ra - 4f_3 \ra^2 - m^2 f_1 \right) a_0}{8f_1 \ra^2}, \\
a_3 & = & -\frac{1}{15 f_1 \ra^2} \left[ \left( 4 f_2 \ra^2 + 20f_1 \ra \right) a_2 + \left( -3 f_3 \ra^2 + 3f_2 \ra + 6f_1 - m^2 f_1 \right) a_1 \right. \nonumber \\
 &&+ \left. \left( - 6 f_4 \ra^2 - 6f_3 \ra +  \frac{\ra^2}{B^2/\mu} f_1^2 - m^2 f_2 \right) a_0 \right], \\
a_4 & = & - \frac{1}{24f_1\ra^2} \left[ \sum_{j=0}^3 (j+2)(2j-4)\ra^2 f_{5-j} a_j  +\sum_{j=0}^3 (j+2)(4j-5)\ra f_{4-j} a_j \right. \nonumber \\
&& + \left. \sum_{j=0}^2 \left( (j+2)(2j-1) -m^2\right) f_{3-j}  a_j  + \sum_{j=0}^1 \sum_{l = 0}^{1-j} \frac{\ra^2}{B^2/\mu}  f_{l+1} f_{2-j-l} a_j + 2\frac{\ra}{B^2 / \mu} f_{1}^2 a_0 \right], \nonumber \\ \\
a_k & = & -\frac{1}{k (k+2)f_1\ra^2} \left[ \sum_{j=0}^{k-1} (j+2) (2j - k) \ra^2 f_{k-j+1} a_j + \sum_{j=0}^{k-1} (j+2) (4j -2 k + 3) \ra f_{k-j} a_j \right. \nonumber \\
&& + \left. \sum_{j=0}^{k-2} \left( (j+2)(2j-k+3) - m^2 \right) f_{k-j-1} a_j +  \sum_{j=0}^{k-3} \sum_{l = 0}^{k-j-3} \frac{\ra^2}{B^2/ \mu} f_{l+1}f_{k-j-l-2} a_j \right. \nonumber \\
&& + \left. \sum_{j=0}^{k-4} \sum_{l = 0}^{k-j-4} \frac{2\ra}{B^2/ \mu} f_{l+1}f_{k-j-l-3} a_j  + \sum_{j=0}^{k-5} \sum_{l = 0}^{k-j-5} \frac{\mu}{B^2} f_{l+1}f_{k-j-l-4} a_j  \right], \quad \textrm{for} \quad k \geq 5, \nonumber \\ \\
s_0 &=& 1, \\
s_1 &=& 0, \\
s_2 &=& 0, \\
s_3 &=& \frac{1}{3f_1\ra^2} \left[ \left(  m^2 f_2 - \frac{\ra^2}{B^2/\mu} f_1^2 \right) s_0 - \mathcal{C} \left( 4\ra^2 f_1 a_1 + (\ra^2 f_2 + 5\ra f_1) a_0 \right) \right] , \\
s_4 &=& -\frac{1}{8f_1\ra^2}  \left[ 9f_1\ra s_3 + \left(\frac{2\ra}{B^2/\mu}f_1^2 +\frac{2\ra^2}{B^2/\mu} f_1 f_2 - m^2 f_3  \right) s_0 \right. \nonumber \\
&& + \left. \mathcal{C} \left( 6\ra^2 f_1 a_2 + 3\ra^2 f_2 a_1 + 9\ra f_1 a_1 + 3(\ra f_2 + f_1)a_0 \right) \right], \\
s_k &=& -\frac{1}{k(k-2)f_1\ra^2} \left\{ \sum_{j=0}^{k-1} j (2j-k-2)\ra^2 f_{k-j+1} s_j + \sum_{j=0}^{k-1} j (4j-2k-1)\ra f_{k-j} s_j \right. \nonumber \\
&& \left. + \sum_{j=0}^{k-2} \left[ \left( j (2j-k+1) - m^2\right) f_{k-j-1} s_j + \mathcal{C} (3j-k+4)\ra^2 f_{k-j-1} a_j \right] \right. \nonumber \\
&& + \left. \sum_{j=0}^{k-3} \left[ \sum_{l=0}^{k-j-3} \frac{\ra^2}{B^2/\mu} f_{l+1} f_{k-j-l-2} s_j + \mathcal{C} (6j-2k+11)\ra f_{k-j-2} a_j \right] \right. \nonumber \\
&& \left. + \sum_{j=0}^{k-4} \left[ \sum_{l=0}^{k-j-4} \frac{2\ra}{B^2/\mu} f_{l+1} f_{k-j-l-3} s_j + \mathcal{C} (3j-k+7) f_{k-j-3} a_j \right] \right. \nonumber \\
&&  + \left. \sum_{j=0}^{k-5} \sum_{l=0}^{k-j-5} \frac{\mu}{B^2} f_{l+1} f_{k-j-l-4} s_j  \right\}, \quad \textrm{for} \quad k \geq 5.
\end{eqnarray}
In the absence of flow, $f_k = \omega^2 \rho_k$ and the coefficients consistently revert to those given in  \citet{paperI}.

\end{appendix}

\end{document}